\documentclass[preprint,amsmath,amssymb]{revtex4}
\usepackage{epsfig}
\usepackage{amssymb}

\usepackage{latexsym}
\usepackage{amsmath}
\usepackage{slashed}
\usepackage{graphics}
\usepackage{epsfig}
\def\br{\begin{eqnarray}}
\def\er{\end{eqnarray}}
\def\be{\begin{equation}}
\def\ee{\end{equation}}

\def\({\left(}
\def\){\right)}
\def\<{\left\langle}
\def\>{\right\rangle}

\begin{document}

\title{The mass splitting in an 331-TC coupled Scenario}

\author{A. Doff}

\email{agomes@utfpr.edu.br}

\affiliation{Universidade Tecnol\'ogica Federal do Paran\'a - UTFPR - DAFIS
Av Monteiro Lobato Km 04, 84016-210, Ponta Grossa, PR, Brazil }

\date{\today}

\begin{abstract}
The root of most of the technicolor (TC) problems lies in the way the ordinary fermions acquire their masses,  where an ordinary fermion (f) couples to a technifermion (F) mediated by an Extended Technicolor (ETC) boson leading  to fermion masses that vary with the ETC mass scale ($M_E$) as $1/M_E^2$.  Recently, we discussed a new approach consisting of models where TC and QCD are coupled through a larger
theory, in this case  the solutions of these equations are modified compared to those of the isolated equations, and TC and QCD self-energies
are of the Irregular form, which allows us to build models where ETC boson masses can be pushed to very high energies. In this  work we extend these results  for 331-TC models, in particular considering a coupled system of Schwinger-Dyson equations, we show that  all technifermions of the model exhibit the same asymptotic behavior for  TC  self-energies. As an application we discuss  how the mass splitting  of the order $O(100)GeV$  could be generated between the second and third generation of fermions. 

\end{abstract}


\maketitle

\section{Introduction}

 \par The gauge and chiral symmetry breaking in quantum field theories can be promoted by fundamental  scalar bosons. However, the main ideas about symmetry breaking and spontaneous generation of fermion and gauge boson masses in field theory were based on the superconductivity theory. Nambu and Jona-Lasinio\cite{Nambu} proposed one of the first field theoretical models where all the most important aspects of symmetry breaking and mass generation were established, as known nowadays. The possibility of spontaneous gauge and chiral symmetry breaking promoted by a composite scalar boson in the context of the  Standard Model (SM) was formulated in the seventies by Weinberg \cite{we} and Susskind \cite{su}. The most popular  version of these models was dubbed as technicolor (TC), where new fermions (or technifermions) condensate and may be  responsible for the chiral and SM gauge symmetry breaking \cite{Rev1,Rev2,Rev3}.

\par  The root of most of the TC problems lies in the way the ordinary fermions acquire their masses,  where an ordinary fermion (f) couples to a technifermion (F) mediated by an Extended Technicolor (ETC) boson. These problems occur when new extended technicolor interactions (ETC) are introduced in order to provide masses to the standard quarks,  leading to quark masses that vary with the ETC mass scale ($M_E$) as $1/M_E^2$. A likely solution of this problem  was proposed by Bob Holdom\cite{holdom} many years ago, remembering that TC  self-energy behaves as $\Sigma_{TC} (p^2)\approx \frac{\left\langle {\bar{F}}F\right\rangle_{\mu}}{p^2} \left(\frac{p}{\mu}\right)^{\gamma_m}$, where $\gamma_m$ is the mass anomalous dimension associated to the fermionic condensate.

\par  Solutions to the above dilemma seem to require a large  $\gamma_m$ value \cite{holdom} leading to a TC self-energy with a harder momentum behavior,  and many models along this idea can be found in the literature \cite{lane0,appel,yamawaki,aoki,appelquist,shro, walk6, walk7, walk8, kura, yama1, yama2,mira2,yama3,mira3,yama4}. In particular we may quote the work of Takeuchi \cite{takeuchi} where the TC Schwinger-Dyson equation (SDE) was solved with the introduction of an \textsl{ad hoc} four-fermion interaction, which can lead to the Irregular solution for the TC self-energy, which is described by Eq.(\ref{eq1})
\be
\Sigma_{TC}(p^2\rightarrow\infty)\propto \mu_{TC}[ 1 + bg^2(\mu^2_{TC})\ln(p^2/\mu^2_{TC})]^{-\delta},
\label{eq1}
\ee
\noindent where $g$ is the TC running coupling constant, $b$ is the coefficient of $g^3$ term in the renormalization group $\beta (g)$ function and
$\delta$ is a function of model parameters. 

\par Recently  we solved numerically the coupled system of the gap equations, TC (based on a $SU(2)$ group) and QCD\cite{us1,us2}. It turned out that both self-energies(TC and QCD) have the same asymptotic behavior of Eq.(\ref{eq1})\footnote{When quarks and technifermions form a coupled system, the overal effect is that the different strong interactions  provide masses  to each other, leading to self-energies behaving as the one shown in Eq.(1)}(with the appropriate exchange in the respective energy scales  $TC \leftrightarrow  QCD$). In this  work we extend the results obtained in\cite{us1,us2} for 331-TC model , whose structure is presented in the sequence. 

\par In particular considering the coupled system described in Fig.(1), we show that only for the coupled system all technifermions present in the model have the same asymptotic behavior given by Eq.(\ref{eq1}). In addition as an application we discuss how the mass splitting between the second and third generation of fermions can be generated in this model.

\begin{figure}[t]
\centering
\vspace{-1cm} 
\includegraphics[scale=0.6]{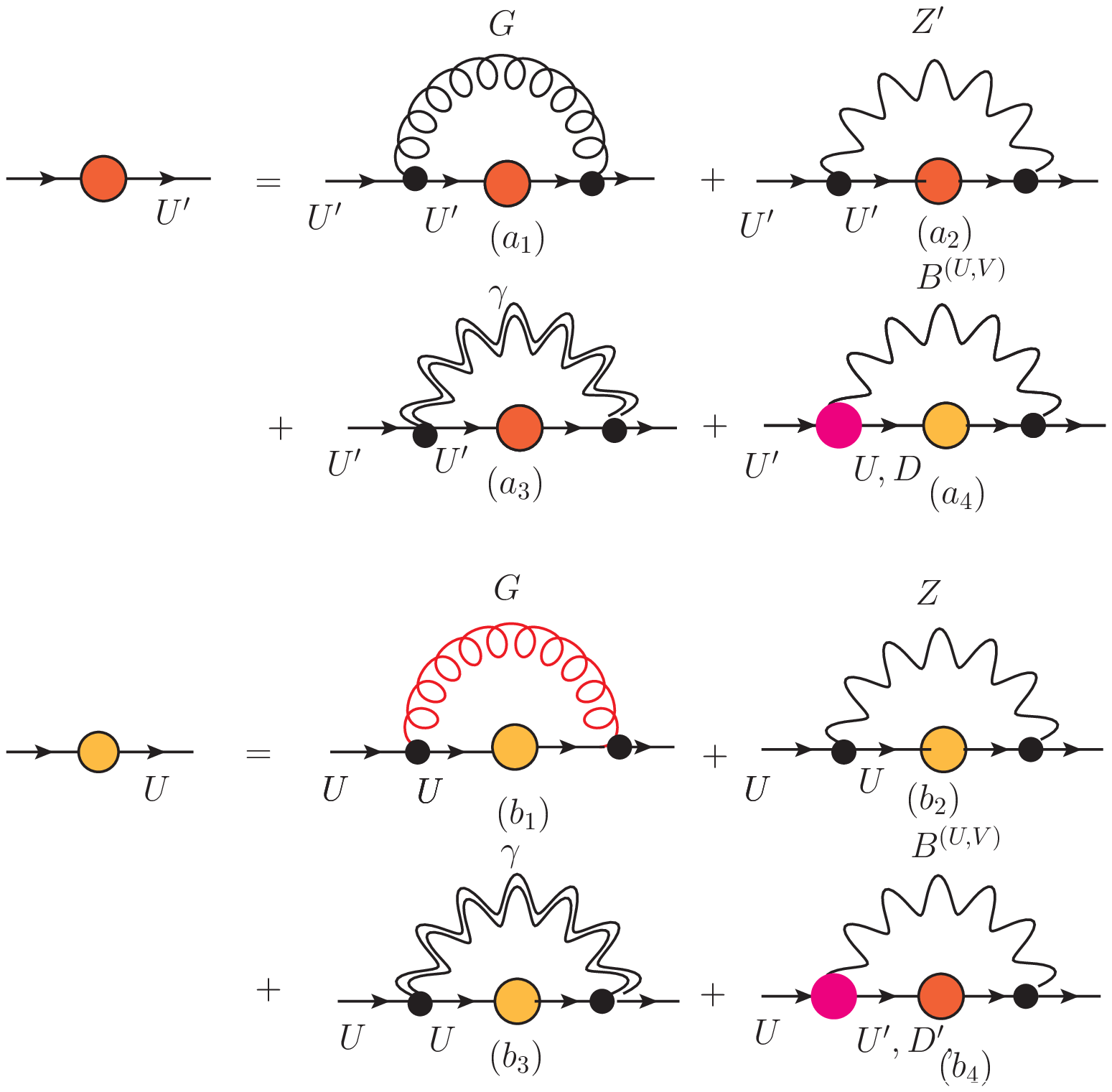} 
\vspace{-7cm} 
\caption[dummy0]{The coupled system of  SDEs for 331-TC: In this figure  (U' $\equiv$ 1)  denotes exotic techniquarks and  (U $\equiv$ 2)  techniquarks  including  ($Z'$, $B^{(U,V)}$) and electroweak corrections. $G$ indicate technigluons interactions.}
\label{fig1}
\end{figure}

\par This  article is organized as follows: In section 2 we present a brief review in relation to the main aspects of 331-TC models, in section 3 we  discuss their coupled Schwinger-Dyson equations (SDE). In section 4  will show how the boundary conditions of the  anharmonic oscillator representation of the coupled gap equation are directly related  with  the mass anomalous dimensions, leading to a $\gamma_m$ to coupled system that is compatible with a self-energy corresponding to Eq.(\ref{eq1}). Finally, in section 5 we discuss  how the mass splitting between the second and third generation of fermions can be generated , in the last section we draw our conclusions.

\section{$SU(N)_{{}_{TC}}\otimes SU(3)_{{}_{L}}\otimes U(1)_{{}_{X}}$ models}

\par  In some  extensions of the standard model (SM), as in the so called 3-3-1 models $SU(3)_{{}_{L}}\otimes SU(3)_{{}_{c}} \otimes U(1)_{{}_{X}}$\cite{331a,331b,331c,331d,331e},  new massive neutral and charged gauge bosons,  $Z'$  and  $ B^{(U,V)} = V^{\pm}, U^{\pm\pm} $, are predicted. The 3-3-1 model is the minimal gauge group that at the leptonic level admits charged fermions and their antiparticles as members of the same multiplet,  the extra predictions of these $GW = SU(2)_{{}_{L}}\otimes U(1)_{{}_{Y}}$ alternative models are leptoquark fermions with electric charges $5/3$ and $-4/3$ and bilepton gauge bosons with lepton number $L = \pm 2$. The quantization of electric charge is inevitable in the $G=3-m-1$(where m=3,4) models\cite{QQ1, QQ2, QQ3, QQ4, QQ5, QQ6} with three non-repetitive fermion generations, breaking universality independently of the character of the neutral fermions.

\par  In the Refs.\cite{df331TC1, Das} it was suggested that the gauge symmetry breaking of a specific version of a 3-3-1 model could be implemented dynamically, because at the scale of a few TeVs the $U(1)_X$ coupling constant becomes strong and the  exotic quark $T$(charge $5/3$)  will form a  $U(1)_X$ condensate breaking  $SU(3)_{{}_{L}}\otimes U(1)_X$ to the electroweak symmetry.

\par This possibility was explored in the Refs.\cite{df331TC2} assuming a model based on the gauge symmetry  $SU(2)_{TC}\otimes SU(3)_{{}_{L}}\otimes SU(3)_{{}_{c}} \otimes U(1)_{{}_{X}}$, where the electroweak symmetry is broken dynamically by a technifermion condensate. In general TC   can be characterized by the $SU(N)_{TC}$ technicolor gauge group, taking the configuration $SU(N)_{{}_{TC}}\otimes SU(3)_{{}_{L}}\otimes U(1)_{{}_{X}}$\cite{df331TC4} that we refer  generically as 331-TC models.
\par On the lines below we describe the main features these models, the fermionic content  has the following form
\br 
&& Q_{3L} = \left(\begin{array}{c} t \\ b \\ T  \end{array}\right)_{L}\,\,\sim\,\,({\bf 1}, {\bf 3}, 2/3) \nonumber \\  \nonumber \\
&&t_{R}\,\sim\,({\bf 1}, {\bf 1}, 2/3)\,,\,b_{R}\,\sim\,({\bf 1},{\bf 1},-1/3)\nonumber  \\  
&&T_{R}\,\sim\, ({\bf 1},{\bf 1},  5/3) \nonumber 
\er
\br
&&Q_{\alpha L} = \left(\begin{array}{c} D \\ u \\ d  \end{array}\right)_{\alpha L}\,\,\sim\,\,({\bf 1}, {\bf 3^*}, -1/3) 
\nonumber \\  \nonumber \\
&&u_{\alpha R}\, \sim\, ({\bf 1},{\bf 1},  2/3 )\,,\,d_{\alpha R}\,\sim\,({\bf 1}, {\bf 1}, -1/3)\nonumber  \\  
&&D_{\alpha R}\,\sim\, ({\bf 1}, {\bf 1},  -4/3 )
\er
\noindent  $\alpha = 1,2 $ is the family index and  we represent  the third quark  family  by $Q_{3L}$. In these expressions $({\bf 1},{\bf 3} ,  X)$, $({\bf 1}, {\bf 3^*},  X)$ or $({\bf 1}, {\bf 1},  X)$ denote the  transformation properties  under  $SU(N_{TC})\otimes SU(3)_{{}_{L}}\otimes U(1)_{{}_{X}}$ and $X  \equiv \frac{Y}{2}$ is the  corresponding $U(1)_{X}$ charge\cite{331b, 331c, Das}.  The leptonic sector includes, besides the  conventional  charged leptons and their respective neutrinos, the charged heavy leptons $E_a$\cite{331e}.  
\br
&& l_{aL} = \left(\begin{array}{c} \nu_{a} \\ l_a \\ E^c_a\end{array}\right)_{L}\,\sim\,({\bf 1},{\bf 3},  0)
\er
\noindent where $a=1,2,3$ is the family index and $l_{aL}$ transforms as triplets   under $SU(3)_L$. Moreover,  we have to add the corresponding right-handed components, $l_{aR} \sim ({\bf 1},{\bf 1}, -1)$ and $E^c_{aR} \sim ({\bf 1},{\bf 1},+1)$.  
\par  Regarding the TC sector the fermionic has the form
\br 
&&\Psi_{1L} = \left(\begin{array}{c} U_1\\ D_1\\  U'\end{array}\right)_{L}\,\,\sim\,\,({\bf N_{TC}}, {\bf 3}, 1/2) \nonumber  \\ 
\nonumber \\ 
&&U_{1 R}\, \sim\, ({\bf  N_{TC}}, {\bf 1}, 1/2)\,,\,D_{1 R}\, \sim \,({\bf  N_{TC}}, {\bf 1},-1/2)\nonumber \\
&&U'_{R}\,\sim\, ({\bf  N_{TC}},{\bf 1}, 3/2)\nonumber  ,
\er
\br 
&&\Psi_{2L} = \left(\begin{array}{c} D'\\ U_2\\  D_2 \end{array}\right)_{L}\,\,\sim\,\,({\bf  N_{TC}}, {\bf 3^*}, -1/2) \nonumber \\ 
\nonumber \\ 
&&U_{2 R}\, \sim\, ({\bf  N_{TC}}, {\bf 1}, 1/2)\,,\,D_{2 R}\,\sim\,({\bf  N_{TC}}, {\bf 1},-1/2)\nonumber \\ 
&&D'_{R}\,\sim\, ({\bf  N_{TC}},{\bf 1}, -3/2), 
\label{3}
\er
\noindent the index $1$ and $2$ label the  first and second techniquark families, $U'$ and $D'$ can be considered as exotic techniquarks making an  analogy with exotic quarks $(T, D)$ that appear in the  ordinary  fermionic content  of the model. The model is anomaly free if we have equal numbers of triplets$(3)$ and antitriplets$(3^*)$, counting the color of $SU(3)_c$. Therefore, in order to make the model anomaly free two of the three quark generations transform as ${\bf 3^*}$,  the third quark  family and the three leptons generations  transform as ${\bf 3}$.  It is easy to check that all gauge anomalies cancel out in this model,  in the TC sector the triangular anomaly cancels between the two generations of technifermions.  In the present version of the model the technifermions are singlets of $SU(3)_c$.


\section{ 331-TC coupled system of Schwinger-Dyson equations}

\par The starting point in this analysis is the diagrammatic representation of the coupled SDEs for the exotic techniquarks $U'$ and
usual techniquarks $U$ self-energies, shown in the first and second lines of Fig. 1, respectively. In this figure the curly lines correspond to  technigluons $(G)$ , the wavy lines to the $(Z, Z', U, V)$  bosons and double wavy lines to fotons $(\gamma)$.
\par  Notice that the above coupled system is rather intricate, it involves different full boson propagators and fully dressed vertices, which should be closely intertwined through the different mass scales of the theories, namely, $(\mu_{331,TC})$. Here, we restrict ourselves to exploring the result of this coupled system in a rather simplified context. Besides that, we neglect the possible contributions that the diagrams $(a_3, b_3)$ may give, since electromagnetic coupling constant $\alpha < \alpha_{X},\alpha_{TC}$. 
\par In the analysis presented in this work, the conventional self-energies of the exotic techniquarks ($a_1$) and techniquarks ($b_1$)  receive the perturbative corrections generated by  $(Z', U, V)$ interactions, and corrections generated by the ETC  discussed in section 5.  The new 3-3-1  bosons $(U,V)$  couples the exotic techniquarks to techniquarks (and vice versa), as represented by the diagrams ($a_4$) and ($b_4$).
\par  Finally, we approximate the fully dressed propagators and vertices, entering into the gap equations, by their tree level expressions. In order to convert the system of coupled equations described in Fig. 1  into a coupled system of differential equations,  we will consider the angle approximation \cite{ang}, transforming  terms as
\be 
\frac{1}{(p -k)^2 + M^2} = \frac{\pi}{2}\left[\frac{\theta(p - k)}{p^2 + M^2} + \frac{\theta(k - p)}{k^2 + M^2}\right]. 
\ee
\noindent where  $M = M_{G,Z'}$ or $M = M_{B^{(U,V)}}$.

\par  Moreover, we defined the notation  indicated by the index $j$, where $(j = 1 = U')$ and $(j = 2 = U)$, leading to the following  TC coupled system of equations 
\br 
&&\Sigma_1(p)= \frac{\theta_1}{p^2 + m^2_G}\int^{p^2}_{0}dk^2\frac{k^2\Sigma_1(k^2)}{k^2+\Sigma^2_1(k^2)} + \theta_1\int^{\Lambda^2}_{p^2} dk^2\frac{k^2\Sigma_1(k^2)}{k^2+\Sigma^2_ 1(k^2)}\frac{1}{k^2 +m^2_G} \nonumber \\
&& + \frac{\theta_{Z'}}{p^2 + M^2_{Z'}} \int^{p^2}_{0} dk^2\frac{k^2\Sigma_1(k^2)}{k^2+\Sigma^2_1(k^2)} + \theta_{Z'}\int^{\Lambda^2}_{p^2} dk^2\frac{k^2\Sigma_1(k^2)}{k^2+\Sigma^2_ 1(k^2)}\frac{1}{k^2 + M^2_{Z'}} \nonumber \\
&& + \frac{\theta_{B_{U,V}}}{p^2 + M^2_{B_{U,V}}} \int^{p^2}_{0} dk^2\frac{k^2\Sigma_2(k^2)}{k^2+\Sigma^2_ 2(k^2)} + \theta_{B_{U,V}}\int^{\Lambda^2}_{p^2} dk^2\frac{k^2\Sigma_2(k^2)}{k^2+\Sigma^2_ 2(k^2)}\frac{1}{k^2 + M^2_{B_{U,V}}}. \nonumber \\
\label{eq1s1}
\er 
\noindent with $\theta_{j} = \frac{3C_{TC}\alpha_{TC}}{4\pi}$, $\theta_{Z'} = \frac{3\alpha_X Y_{{}_{L}}Y_{{}_{R}}}{4\pi}$ and $\theta_{B_{V,U}} = \frac{3\alpha_L Y_{{}_{L}}Y_{{}_{R}}}{4\pi\alpha_{TC}\sqrt{2}}$. In this expressions the coupling constants $\alpha_L = g^2_{331}/4\pi$, $\alpha_X = g^2_X/4\pi$  are associated to the gauge group $SU(3)_{{}_{L}}\otimes U(1)_{{}_{X}}$ , and 
\br
\frac{\alpha_{{}_{X}}}{\alpha_L} =  \frac{\sin^2\theta_{{}_{W}}(\mu)}{1 - 4\sin^2\theta_{{}_{W}}(\mu)}
\label{eqU1}
\er
\noindent  $\theta{{}_{W}}$ is the electroweak mixing angle and $\mu=\mu_{331}$ is 331  mass scale$(\sim O(TeV))$. 

\par The effective coupling indicated by $\theta_{B_{U,V}}$(Magenta bubble in Fig.(1)) was determined in Ref.\cite{df331TC6}. For the techniquark $U$ self-energy we have a similar equation just changing the index  $1 \leftrightarrow 2$, with $Z' \to Z$ and $\theta_{Z} = \frac{3\alpha_{w}C_{w}}{4\pi}[w=weak]$.
\par  In order to obtain the gap equation, Eq.(\ref{eq1s1}), in the dimensionless form we will consider the definitions below
\br 
&& p^2 = M^2_{j}x \,\,\,,\,\,\, \Sigma_j (p^2) = M_j f_j(x)\,\,\,,\,\,\, \delta_{G} =\frac{m^2_G}{M^2_j}\nonumber \\
&& k^2 = M^2_{j}y \,\,\,,\,\,\,\Sigma_j (k^2) = M_j f_j(y)\,\,\,,\,\,\,\delta_{Z, Z', B_{U,V}} =\frac{M^2_{Z, Z',B_{U,V}}}{M^2_j}\nonumber \\
\label{eq1s2}
\er 
\noindent  where  $M_j = M_j(0)$   correspond respectively to the dynamical  masses  ,  $m_G$ represents technigluons dynamical mass scale \cite{mg1,mg2,mg3,mg4}, whereas $M_Z$ corresponds to the $Z_0$  mass  and $M_{Z',B_{U,V}}$ gauge boson masses  related to  $SU(3)_{{}_{L}}\otimes U(1)_{{}_{X}}$ model. Therefore, considering the set of parameters identified above, we obtain    
\br 
&&f_1(x)=\frac{\theta_1}{x + \delta_G}\int^{x}_{0}dy\frac{yf_1(y)}{y+f^2_1(y)} +  \theta_1\int^{\frac{\Lambda^2}{M^2_1}}_{x} dy\frac{yf_1(y)}{y+f^2_ 1(y)}\frac{1}{y +\delta_G} \nonumber \\
&& + \frac{\theta_{Z'}}{x + \delta_{Z'}}\int^{x}_{0} dy\frac{yf_1(y)}{y+f^2_1(y)}  + \theta_{Z'}\int^{\frac{\Lambda^2}{M^2_1}}_{x} dy\frac{yf_1(y)}{y+f^2_1(y)}\frac{1}{y + \delta_{Z'}} \nonumber\\  
&& + \frac{\theta_{B_{U,V}}}{x + \delta_{\theta_{B_{U,V}}}}\left(\frac{M_2}{M_1}\right) \int^{x}_{0} dy\frac{yf_2y)}{y+\frac{M^2_2}{M^2_1}f^2_2(y)}  \nonumber\\ 
&& + \theta_{B_{U,V}}\left(\frac{M_2}{M_1}\right)\int^{\frac{\Lambda^2}{M^2_1}}_{x} dy\frac{yf_2(y)}{y+\frac{M^2_2}{M^2_1}f^2_ 2(y)}\frac{1}{y + \delta_{\theta_{B_{U,V}}}} , 
\label{eq1s3}
\er
\noindent  so that for the coupled system the integral equation above, together with the equation for $j=2$, lead to
\br 
&& f_1(x)= \sigma_1(x)I^a_1(x) + \theta_1(x)I^b_1(x) + \sigma_{Z'}(x)I^a_1(x) + \theta_{Z'} I^b_{Z'}(x)  \nonumber \\
&& \hspace*{1.1cm} + \, \sigma_{B_{U,V}}(x)\left(\frac{M_2}{M_1}\right) I^a_{2\frac{M_2}{M_1}}(x) +  \theta_{B_{U,V}}\left(\frac{M_2}{M_1}\right) I^b_{2 B_{U,V}}(x) \nonumber \\
&& f_2(x)= \sigma_2(x)I^a_2(x) + \theta_2(x)I^b_2(x) + \sigma_{Z}(x)I^a_2(x) + \theta_{Z} I^b_{Z}(x)  \nonumber \\
&& \hspace*{1.1cm} + \, \sigma_{B_{U,V}}(x)\left(\frac{M_1}{M_2}\right) I^a_{1\frac{M_1}{M_2}}(x) +  \theta_{B_{U,V}}\left(\frac{M_1}{M_2}\right) I^b_{1 B_{U,V}}(x). 
\label{eq1s5}
\er 
\noindent The variables introduced in Eq.(\ref{eq1s5}), for $j,k = 1,2$ and $k \neq j$, are described in the sequence
\br 
&&\sigma_j(x) = \frac{\theta_j}{x + \delta_G}\,\,\,,\,\,\,\sigma_{Z,Z',B_{V,U}}(x) = \frac{\theta_{Z, Z', B_{V,U}}}{x + \delta_{Z, Z', B_{V,U}}}\nonumber \\
&& I^a_j(x) = \int^{x}_{0}dy\frac{yf_j(y)}{y+f^2_j(y)} \,\,\,,\,\,\, I^b_j(x) = \int^{\frac{\Lambda^2}{M^2_j}}_{x} dy\frac{yf_j(y)}{y+f^2_ j(y)}\frac{1}{y +\delta_G} \nonumber \\
&&  I^b_{Z,Z'}(x) = \int^{\frac{\Lambda^2}{M^2_j}}_{x} dy\frac{yf_j(y)}{y+f^2_ j(y)}\frac{1}{y + \delta_{Z,Z'}}\,\, ,\,\,  I^a_{k\frac{M_j}{M_k}}(x)= \int^{x}_{0}dy\frac{yf_j(y)}{y + \frac{M^2_j}{M^2_k}f^2_j(y)}\nonumber \\
&&  I^b_{j B_{U,V}}(x)  = \int^{\frac{\Lambda^2}{M^2_k}}_{x} dy\frac{yf_j(y)}{y+\frac{M^2_j}{M^2_k}f^2_ j(y)}\frac{1}{y + \delta_{B_{U,V}}},
\er 

\noindent for  $M_1 = M_2$, the coupled equation system becomes equivalent to a system composed of identical particles that
  are coupled by the term $\theta_{B_{U,V}}$, and in this case it is possible to assume the following simplifications 
\br 
&& \sigma_1(x) = \sigma_2(x)   \nonumber \\
&& I^a_1(x) = I^a_2(x) = I^a_{1\frac{M_2}{M_1}}(x)= I^a_{2\frac{M_1}{M_2}}(x)=I^a(x).  
\label{eq1s6}
\er 
\par The system  of the integral equations above, Eq.(\ref{eq1s5}), now can be converted into a differential equation system. As  consequence, the derivatives $f'_j(x)$ of Eq.(\ref{eq1s5}) present a series of cancellations between the terms proportional to $I^a(x)$, and $I^b_j(x)$,$I^b_{Z,Z',B_{U,V}}(x)$ leading to the following set of  equations 

\br 
&& f_1''(x) + H_1(x)f_1'(x) +  P_1(x)f_1(x) = 0 \nonumber \\
&& f_2''(x) + H_2(x)f_2'(x) +  P_2(x)f_2(x) = 0,
\label{eq1s6}
\er 

\noindent where  
\br 
&& H_j(x)\equiv 2\frac{(\frac{\sigma_j(x)}{(x + \delta_G)^2} + \frac{\sigma_{Z',Z}(x)}{(x + \delta_{Z',Z})^2} +\frac{\sigma_{B_{U,V}}(x)}{(x + \delta_{B_{U,V}})^2})}{(\frac{\sigma_j(x)}{(x + \delta_G)} + \frac{\sigma_{Z',Z}(x)}{(x + \delta_{Z',Z})}+ \frac{\sigma_{B_{U,V}}(x)}{(x + \delta_{B_{U,V}})})}\nonumber \\
&& P_j(x) \equiv  (\frac{\sigma_j(x)}{(x + \delta_G)} + \frac{\sigma_{Z',Z}(x)}{(x + \delta_{Z',Z})} +\frac{\sigma_{B_{U,V}}(x)}{(x + \delta_{B_{U,V}})}).
\er
\par The function $H_j(x)$ can be rewritten as

\be 
 H_j(x)\equiv \frac{2}{(x + \delta_G)}\left[\frac{1  + \kappa_{Z',Z}\frac{(x + \delta_{G})^3}{(x + \delta_{Z',Z})^3} + \kappa_{B_{U,V}}\frac{(x + \delta_{G})^3}{(x + \delta_{Z',Z})^3}}{1 + \kappa_{Z',Z}\frac{(x + \delta_{G})^2}{(x + \delta_{Z',Z})^2} + \kappa_{B_{U,V}}\frac{(x + \delta_{G})^2}{(x + \delta_{Z',Z})^2} }\right],
\label{eqH}
\ee
\noindent in the last expression we identify  $\kappa_{Z',Z} = \frac{\theta_{Z',Z}}{\theta_j}$ and $\kappa_{B_{U,V}} = \frac{\theta_{B_{U,V}}}{\theta_j}$. In order to obtain the asymptotic behavior for Eq.(\ref{eq1s6}), we can consider the limit $x = \delta_{Z'} \approx  \delta_{B_{U,V}} \to  \infty $, which implies 
\br 
&&\hspace*{-0.5cm}H_1(x) \approx \frac{2}{(x + \delta_G)}\left[\frac{1  + \frac{\kappa_{1}}{8}}{1 + \frac{\kappa_{1}}{4}}\right]\,\,,\,\,H_2(x) \approx \frac{2}{(x + \delta_G)}\left[\frac{1  + \kappa_{2}}{1 + \kappa_{2} + \frac{\kappa_{B_{U,V}}}{8}}\right]
\er 
\noindent being the effective couplings $\kappa_{j}$, listed below
\br
&& \kappa_1= \kappa_{Z'} + \kappa_{B_{U,V}}\nonumber \\
&& \kappa_2=  \kappa_{Z} + \frac{\kappa_{B_{U,V}}}{8}. 
\label{eq16}
\er
\noindent As consequence, the Eq.(\ref{eq1s6}) in the  asymptotic region  can be represented by
\br 
&&\hspace*{-1cm} f_1''(x) + \frac{2}{(x + \delta_G)}\left[\frac{1  + \frac{\kappa_{1}}{8}}{1 + \frac{\kappa_{1}}{4}}\right]f'_1(x) +  \frac{\theta_1}{(x + \delta_G)^2}(1 + \frac{\kappa_{1}}{4}) \frac{xf_1(x)}{x+f^2_1(x)} = 0 \nonumber \\ 
&&\hspace*{-1cm} f_2''(x) + \frac{2}{(x + \delta_G)}\left[\frac{1  + \kappa_{2}}{1 + K_2 }\right]f'_2(x) +  \frac{\theta_1}{(x + \delta_G)^2}(1 + K_2) \frac{xf_2(x)}{x+f^2_2(x)} = 0,
\label{eq1s8}
\er 
\noindent where in the last expression  $ K_2 \equiv \kappa_{2} + \frac{B_{U,V}}{8}$. Note that the importance of transforming the coupled system of integral equations in a coupled system of differential equations is that we can now establish a connection with a very interesting analysis of the problem of dynamical symmetry breaking made by Cohen and Georgi\cite{georgi}, who reduced this problem to the one of a mass subjected to an anharmonic potential.

\par  The representation for the TC  gap equation discussed in Ref.\cite{georgi}  corresponds to the  equation of a unit  mass subjected to the anharmonic potential    
\br
&& V(X) = \frac{1}{2}\left[3X^2(t) + a\ln\left( 1 + X^2(t) \right)\right],
\label{eq1s9}
\er
\noindent which is  quadratic  with  a  logarithmic correction  due  to the $SU(N)_{TC}$ gauge theory, where  the differential  form of this  gap equation  is given by
\be 
\ddot{X}(t) + 4\dot{X}(t) + 3X(t) + a\frac{X(t)}{1 + X^2(t)}=0,
\label{eq1s10} 
\ee 
\noindent  with $a = \frac{\theta}{\theta_c} = \frac{\alpha_{TC}}{\alpha_{TC(c)}} = \frac{\alpha}{\alpha_{c}}$ and $t=\frac{1}{2}\ln(\frac{p^2}{\mu^2_{TC}}) $.

\par For the system of equations discussed throughout this section ,  described by Eq.(\ref{eq1s8}), $SU(3)_{{}_{L}}\otimes U(1)_{{}_{X}}$ radiative corrections will lead to corrections to this potential  that can be parametrized by the terms $\kappa_{Z',B_{U,V}}$. In the next section, we will determine how such corrections will lead to $\gamma_m$  compatible with  a self-energy equal to Eq.(\ref{eq1}).


\section{331-TC coupled system in the anharmonic oscillator representation}

\par In order to determine the corresponding Eq.(\ref{eq1s9}) of our problem, it is necessary to introduce the following new variables

\br
&& t=\frac{1}{2}\ln(x) \,\,\,, \,\,\,f_j(x) = e^{\frac{t}{2}} X_j(t)\nonumber \\ 
&& \delta_1(\kappa_1) = \frac{\kappa_1}{8 + 2\kappa_1}\,\,\,, \,\,\,\delta_2(\kappa_2) = \frac{\frac{\kappa_{B_{U,V}}}{8}}{1 + K_2}
\label{eq1s11}
\er

\noindent which take the following form for the  gap equation  system, Eq.(\ref{eq1s8}),  in the representation introduced in Ref.\cite{georgi}
\br 
&&\ddot{X_j}(t) + 4(1 - \delta_j(\kappa_j)) \dot{X}_j(t) + 3X_j(t) + a'_j\frac{X_j(t)}{1 + X^2_j(t)} = 0 ,
\label{eq1s12} 
\er
 \noindent where we define $ a'_j \equiv a  + \theta_j \kappa_j  - 4\delta_j(\kappa_j) $.   The potential described by Eq.(\ref{eq1s9}) is now corrected by $a \to a'_j$,  with $ j = 1,2 $ that contain the corrections resulting from  of Eqs.(\ref{eq1s8}) described in the previous section.

\par  In the limit of small and large  $X(t)$  the  potential of Eq.(\ref{eq1s9})  is  approximately  harmonic, and in these limits the criticallity condition of Eq.(\ref{eq1s10}) can be analyzed, making analogy with the critical behavior shown by a damped harmonic oscillator subjected to the boundary conditions in the infrared (IR)[$t=t_0$] and ultraviolet (UV)[$t=t_{\Lambda}$]  
regions\cite{georgi,Richard} 
\br 
&& \lim_{t\rightarrow t_0} \frac{\dot{X}(t)}{X(t)} = -1 \nonumber \\
&& \lim_{t\rightarrow t_{\Lambda}} \frac{\dot{X}(t)}{X(t)} = -3, 
\label{eq2s1} 
\er 
\noindent where $t_0 = \ln\frac{p}{\mu_{TC}}(p\to 0)$  and  $t_{\Lambda} = \ln\frac{\Lambda}{\mu_{TC}}(\Lambda\to \infty)$. 
\par Eq.(\ref{eq1s10}) is described by a damped harmonic oscillator  in the limit of small $X(t)$, which corresponds to the known behavior of the gap equation solution in the asymptotic region, $t \to t_{\Lambda}$, obtained for $a <1 $. According to Ref.\cite{georgi} precisely in this case OPE(Operator Product Expansion)  provides an interpretation  of  the  parameters  appearing  in  the asymptotic  solution of the  gap  equation.
\par The solution of the corresponding linearized equation [Eq.(\ref{eq1s10})] for $ a < 1$ is described by
\be 
 X(t)  = Ae^{-(2  + \sqrt{1-a})t} + Be^{-(2  - \sqrt{1-a})t} = Ae^{-(3-\gamma_m)t} + Be^{-(1+\gamma_m)t} .
\label{eq2s2}     
\ee 
\noindent where the  mass anomalous dimension ($\gamma_m$) of the techniquark condensate $\langle \bar{Q} Q\rangle$, can be identified as $\gamma_m = 1 - \omega = 1-\sqrt{1-a}$. Moreover, dynamical chiral symmetry breaking  does not occur for $a <1$, on the other hand it is possible to investigate the critical behavior of this gap equation when $a \to 1$ with the following  transformation \cite{georgi}
\be 
Y(t) = e^{(1+\gamma_m)t}X(t)
\label{eq2s3} 
\ee
 
\noindent with this new coordinate shown in Eq.(\ref{eq2s3}) we can verify the following relation between  $X(t)$ and $Y(t)$
\be 
\frac{\dot{X}(t)}{X(t)} = -(1+\gamma_m) + \frac{\dot{Y}(t)}{Y(t)}. 
\label{eq2s3b} 
\ee 
\par Now the differential equation satisfied by $Y(t)$  takes the form\cite{georgi}
\be
\ddot{Y}(t) + 2\sqrt{1-a}\dot{Y}(t) -a\frac{Y^3(t)}{Y^2(t) + e^{2(1+\gamma_m)t}}=0
\label{eq2s4} 
\ee 
\noindent  and  the boundary conditions for  $Y(t)$ in  the infrared (IR)[$t=t_o$] and ultraviolet (UV)[$t=t_{\Lambda}$]  regions can be obtained from Eq.(\ref{eq2s1}), leading to
\br 
&& \lim_{t\rightarrow t_0} \frac{\dot{Y}(t)}{Y(t)} = \gamma_m \nonumber \\
&& \lim_{t\rightarrow t_{\Lambda}} \frac{\dot{Y}(t)}{Y(t)} = \gamma_m -2. 
\label{eq2s5} 
\er 
\par As can be verified  the gap equation for the TC coupled system in  the anharmonic oscillator representation are modified by  $SU(3)_{{}_{L}}\otimes U(1)_{{}_{X}}$   interactions. The solution of the corresponding linearized equation, Eq.(\ref{eq1s12}),  now for $ a'_j < 1$ is described by
\br 
&&\hspace*{-2.5cm} X(t)  = Ae^{-(2 - 2\delta_j(\kappa_j) + \sqrt{1-a'_j})t} + Be^{-(2 - 2\delta_j(\kappa_j) - \sqrt{1-a'_j})t} = Ae^{-(3-\gamma'_{jm})t} + Be^{-(1+\gamma'_{jm})t} 
\label{eq2s6}     
\er 
\noindent whereas for the coupled system the  mass anomalous dimension ($\gamma'_{jm}$) of the techniquark condensate $\langle \bar{Q} Q\rangle^j$  is 
 now changed to $\gamma'_{jm} = \gamma_m + 2\delta_j(\kappa_j)$.

\par Therefore, the boundary condition for  $Y(t)$ in  the  ultraviolet (UV)[$t=t_{\Lambda}$]  region for the coupled system, 
can be obtained from Eq.(\ref{eq2s5}), with the replacement $\gamma_m \to \gamma'_{jm}$ leading to
\br 
&& \lim_{t\rightarrow t_{\Lambda}} \left(\frac{\dot{Y}(t)}{Y(t)}\right)_j = \gamma'_{jm} -2 = \gamma_{m} + 2\delta_j(\kappa_j) - 2. 
\label{eq2s7} 
\er 
\noindent   The Eq.(\ref{eq2s7}) shows  that we will have a much smoother dependency than $1/p$ (walk) at $\gamma_m=1$  limit, since we will have for the coupled system $\left(\frac{\dot{Y}(t_{\Lambda})}{Y(t_{\Lambda})}\right)_j > -1$. Besides that,  the above expression naturally recovers the Lane condition, \cite{Lane} $(2\delta < 1 )$, where $\delta$ represents the total correction to $\gamma_{m}$.

\section{The mass splitting between the second and third generation of fermions}

\par  The mass generated for ordinary fermions in TC models appears through the diagram shown in Fig.(2), where an ordinary fermion (f) couples to a technifermion $(F)$ mediated by an Extended Technicolor (ETC) boson.
\begin{figure}[h]
\centering
\vspace*{-0.5cm}
\includegraphics[scale=0.6]{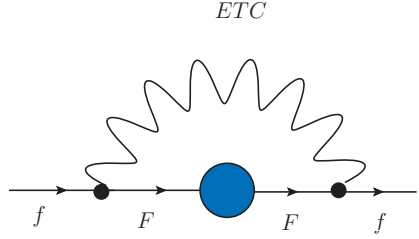} 
\vspace{-0.25cm} 
\caption[dummy0]{Ordinary fermion mass (f) in ETC models}
\label{fig2}
\end{figure}
\par As consequence, it is possible  verify that the mass generated for the fermion (f) is given by\cite{holdom} 
\br 
&& m_f \approx C_{ETC}\alpha_{ETC}\frac{\mu^3_{{}_{TC}}}{M^2_{ETC}}\left(\frac{M_{ETC}}{\mu_{{}_{TC}}}\right)^{\gamma_m}. 
\label{eqmf}
\er
\par  In order to consider the contributions of radiative corrections for coupled system described in Fig.(1), without specifying a model, we will
assume that  $(Q', Q, q)$ and TC are embedded into a ETC gauge group, where $Q=(U,D)$ , $Q'=(U',D')$  and $q = (f,f')$. The  mass anomalous dimension  $\gamma_m$  indicated in Eq.(\ref{eqmf}) should be replaced by $\gamma'_{jm}$ with $F=(Q,Q')$. In this case the calculation for $(m_f)$ obtained from Fig. (2), leads to
\br
&& m_{fj}  \sim C_{ETC}\alpha_{ETC}\frac{\mu^3_{{}_{TC}}}{M^2_{ETC}}\left(\frac{M_{ETC}}{\mu_{{}_{TC}}}\right)^{\gamma_m + 2\delta_j(\kappa_j)}. 
\er
\par  In addition to the corrections shown in Fig. 1, radiative ETC corrections  to  $\gamma_m$ lead to $\gamma_m  + 2\delta_{E}(\kappa_{E})$, with 
$$
\delta_{E}(\kappa_{E}) = \frac{\kappa_E}{8 + 2\kappa_E}
$$
\noindent and  $\kappa_E \approx \alpha_E$.  For a walking theory , where $\gamma_m \approx 1$, we can finally estimate that 
\br 
&& m_{fj} \sim  C_{ETC}\alpha_{ETC}\frac{\mu^2_{{}_{TC}}}{M_{ETC}}\left(\frac{M_{ETC}}{\mu_{{}_{TC}}}\right)^{ 2\delta_j(\kappa_j) + 2\delta_{E}(\kappa_{E}) } = c\left(\frac{M_{ETC}}{\mu_{{}_{TC}}}\right)^{2\delta_j(\kappa_j) + 2\delta_{E}(\kappa_{E}) - 1}  
\label{eqgamma}
\er
\noindent  where  $c =  C_{ETC}\alpha_{ETC}\mu_{{}_{TC}}$. Considering the  Eqs.(\ref{3}), (\ref{eq16}) and (\ref{eq1s11}), assuming $g^2_{X} \approx g^2_{{}_{TC}}$ 
and also the MAC hypothesis $C_{TC}\alpha_{TC} \sim 1$\cite{mac}, with $\alpha_E \sim 0.032$\cite{us1,us2}, we get  $2\delta_1(\kappa_1)  + 2\delta_{E}(\kappa_{E}) \sim O(0.65)$. A similar estimate considering Eqs.(\ref{3}), (\ref{eqU1}), (\ref{eq16}) and (\ref{eq1s11}) can be made for $j=2$ leading to $2\delta_2(\kappa_2) + 2\delta_{E}(\kappa_{E}) \sim O(0.3)$.

\par In the Ref.\cite{dne} we consider a mechanism to generate mass split between  different  femionic generations,  the first fermionic generation basically obtain mass  due  to the interaction with the QCD condensate, characterized by the effective scalar boson $\eta$. Whereas the third generation obtain mass due to its coupling with an TC condensate characterized by the effective scalar boson $\varphi$. The reason for this particular coupling is provided by the $SU(3)_H$ horizontal symmetry, where we assign different  horizontal quantum numbers to first and third fermionic generations  such that the couplings between these generations happen with the composite scalar bosons $\eta$ and $\varphi$. 

\par  The condensates formed by $Q$ and $Q’$ can be related to a system formed by two composite scalars denote by $\phi$ and $\phi’$. It is possible to consider a similar mechanism to that discussed in Ref.\cite{dne},  where the introduction of a horizontal symmetry would provide only the coupling between the effective scalars $\phi$ and $ \phi'$ with fermions of the second (f) and third generation (f') so that  fermion couplings with exotic techniquarks $(U', D')$ are represented by $(f')$, while with usual techniquarks by $(f)$.

\par In Fig. (3) we show the behavior of Eq.(\ref{eqgamma}) for $SU(2)_{TC}$ , the curve (dotted-blue)  correspond to  $ m_{f'}$ , while the (dashed-orange)  to $m_{f}$. 

\begin{figure}[t]
\centering
\vspace*{-0.2cm}
\includegraphics[scale=1.1]{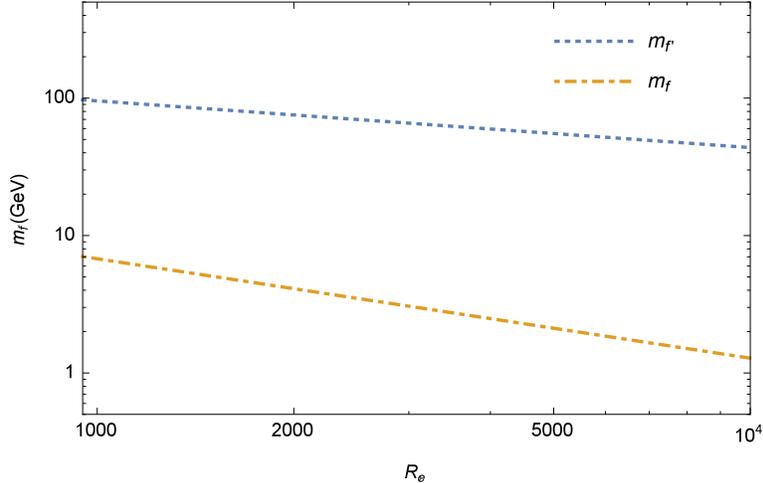} 
\vspace{-0.25cm} 
\caption[dummy0]{This figure shows the behavior of Eq.(\ref{eqgamma}), where $m =m_{fj}$ and  $R_e = \left(\frac{M_{ETC}}{\mu_{{}_{TC}}}\right)$. 
The curves depicted  were obtained for technifermions in the fundamental representation of $SU(N_{TC})$[R=F].}
\label{fig4}
\end{figure}


\section{Conclusions}

\par  In the Refs.\cite{us1,us2} we discussed a new approach consisting of models where TC and QCD are coupled through a larger
theory, in this case  the solutions of these equations are modified compared to those of the isolated equations 
, which allows us to build models where ETC boson masses can be pushed to very high energies.  We calculated numerically the self-energy ($\Sigma (p^2)$) of two coupled strong interaction theories and  verified that the coupled TC self-energy behaves as Eq.(\ref{eq1}). 

\par   In this  work we extend the results obtained in\cite{us1,us2} for 331-TC models ,  considering the coupled system described in Fig.(1)
we find that $(U',U)$ self-energies have a much smoother dependence with the momentum than $1/p$ (walk). The  diagrams on the right-hand side of Fig.(1), $(a_4)$ and $(b_4)$,  plays the same role of the  ETC interactions in the coupled system discussed in \cite{us1,us2}. The  $V^{\pm} ,U^{\pm\pm}$ interactions  modify the UV boundary condition of the SDE in differential form exactly as happens in (QCD and $SU(2)_{{}_{TC}}$) case in such a way that the ultraviolet behavior of the $(U',U)$self-energies  turn out to be of the form of Eq.(\ref{eq1}),  being this result confirmed by the behavior presented in Eq.(\ref{eqgamma}).

\par  The condensates formed by $Q$ and $Q’$ can be related to a system formed by composite scalars $\phi(\phi’)$, in the section 5 considering an extension of the mechanism proposed in Ref.\cite{dne}, we discussed the possibility of generating the  mass splitting between the second and third generation of fermions.
The Fig.(3) illustrates how the mass splitting  of the order $O(100) GeV$(dashed-dotted lines)  could be generated for $\frac{M_{ETC}}{\mu_{{}_{TC}}} = 10^{3}$ or $ M_{ETC} = O(10^{6}) GeV$ for $ c \sim  O(1)TeV$. 

\par The composite scalars bosons $\phi(\phi')$ mimics the behavior of a (2HDM)  that can lead to interesting phenomenological implications. The few aspects discussed here show that models along  the line of coupled strong interactions may open way for the construction of realistic theories of dynamical symmetry breaking.

\section*{Acknowledgments}

This research  was  partially supported by the Conselho Nacional de Desenvolvimento Cient\'{\i}fico e Tecnol\'ogico (CNPq)
under the grant 302663/2016-9.

\end{document}